# AutoPET III Challenge: PET/CT Semantic Segmentation

Reza Safdari[1], Mohammad Koohi-Moghaddam[2], Kyongtae Tyler Bae[3]

[1] safdari@hku.hk, The University of Hong Kong

[2] koohi@hku.hk, The University of Hong Kong

[3] baekt@hku.hk, The University of Hong Kong

## Abstract

In this study, we implemented a two-stage deep learning-based approach to segment lesions in PET/CT images for the AutoPET III challenge. The first stage utilized a DynUNet model for coarse segmentation, identifying broad regions of interest. The second stage refined this segmentation using an ensemble of SwinUNETR, SegResNet, and UNet models. Preprocessing involved resampling images to a common resolution and normalization, while data augmentation techniques such as affine transformations and intensity adjustments were applied to enhance model generalization. The dataset was split into 80% training and 20% validation, excluding healthy cases. This method leverages multi-stage segmentation and model ensembling to achieve precise lesion segmentation, aiming to improve robustness and overall performance.

Keywords: PET/CT tumor lesions segmentation, Ensemble Learning

## Introduction

Cancer is globally recognized as one of the leading causes of premature mortality, alongside cardiovascular diseases, and poses a significant global health challenge. Early detection of cancer lesions is crucial for improving survival rates, as the prognosis and treatment options depend on the location and stage of the lesions.

Computed tomography (CT) and positron emission tomography (PET) are essential for tumor analysis, providing vital information about the tumor's location, anatomy, and stage. However, manual analysis of medical images is time-consuming and error-prone, leading to inconsistencies among specialists. Therefore, there is a pressing need to develop and integrate deep learning methods into healthcare to enhance diagnostic processes and better understand cancer mechanisms.

With this in mind, we participated in the AutoPET-III challenge held in MICCAI 2024, which aims to refine automated segmentation of tumor lesions in Positron Emission Tomography/Computed Tomography (PET/CT) scans in a multitracer multicenter setting.

The goal of the competition is to develop models that can accurately segment FDG- and PSMA-avid tumor lesions in whole-body PET/CT images, while avoiding false-positive segmentation of anatomical structures with physiologically high uptake. Additionally, the challenge focuses on building models that generalize well across different tracers, acquisition protocols, and clinical sites, addressing the critical need for robust and efficient automated analysis in oncological diagnostics.

In this study, a two-stage deep learning-based approach was implemented to segment lesions in PET/CT images in the AutoPET-III challenge [1-3]. Two distinct stages were employed, utilizing different neural network architectures for each stage, with an ensemble of models used in the second stage for final segmentation.

## Methodology

In this study, a two-stage deep learning approach was developed for lesion segmentation in PET/CT images as part of the AutoPET-III challenge. The method involved using different neural network architectures for each stage, with an ensemble of models applied in the second stage to produce the final segmentation, benefiting from both coarse-to-fine segmentation refinement and the diverse strengths of multiple deep learning architectures.

Stage 1: Preliminary Segmentation

For the first stage, a **DynUNet** [4] model was used to perform a coarse segmentation of the lesion areas. The input images were preprocessed into patches of size **(128, 160, 112)**. This stage aims to identify broad regions of interest (ROIs) where lesions may be present, providing a rough segmentation that is refined in the second stage.

Stage 2: Fine Segmentation with Ensemble

In the second stage, the results from the first stage were used as additional input to refine the lesion segmentation further. Before feeding the first-stage segmentation masks into the second-stage models, **random coarse dropout** was applied to these masks to increase robustness and prevent over-reliance on the initial predictions. The second stage employed an ensemble of three models:

- **SwinUNETR** [5]: A transformer-based model with a patch size of **(96, 96, 96)**, designed to capture global and long-range dependencies.
- **SegResNet** [6]: A residual network-based architecture, trained on patches of size **(192, 192, 192)**, to capture fine details in the lesion segmentation.
- **UNet** [7]: A classic U-Net architecture, trained with a patch size of **(128, 160, 112)**, to further refine the segmentation results from the first stage.

The final segmentation results were obtained by **ensembling** the outputs of these three models, aggregating their predictions to improve overall segmentation performance.

# Results

## Dataset

The dataset for the AutoPET-III challenge comprises two cohorts: FDG and PSMA. The FDG cohort includes 501 patients with malignant melanoma, lymphoma, or lung cancer, and 513 negative controls, while the PSMA cohort contains PET/CT images from 537 patients with prostate carcinoma, with and without PSMA-avid lesions. The training datasets show different age distributions, with FDG patients being younger (mean age: 58-60) compared to PSMA patients (mean age: 71). Imaging conditions also varied, with the FDG data acquired using a single scanner, while the PSMA data was collected across three different PET/CT scanners.

The PET/CT acquisition protocols differ between the FDG and PSMA datasets. FDG images were obtained after a six-hour fasting period and one hour post-injection of 350 MBq 18F-FDG. PSMA images were acquired approximately 74 minutes after injecting 18F-PSMA or 68Ga-PSMA. Annotation for both datasets involved radiologists with 5-10 years of experience, following a two-step protocol: the identification of tracer-avid lesions and manual segmentation of the lesions in axial slices. A new version, Dataset v1.1, was released after discovering three cases that did not meet the quality criteria were mistakenly included.

The dataset was randomly split into 80% training and 20% validation without k-fold cross-validation. Images without lesions (healthy cases) were excluded from both training and validation to focus the model on cancerous cases. Additionally, only samples from version 1.1 of the competition dataset were used, with retracted samples being excluded from the analysis.

## Experimental Settings

All volumes were resampled to a uniform resolution by adjusting each image's spacing to match the dataset's average spacing. Following resampling, images were normalized to maintain a consistent intensity distribution throughout the dataset. To enhance model generalization, various augmentations were applied during training for both stages, including random affine transformations (translation, rotation, scaling), random Gaussian noise and smoothing, random intensity and contrast adjustments, and random flipping. All models. All methods were implemented using the MONAI framework. We trained the models for 774 epochs with a learning rate of 1e-3 and a weight decay of 3e-5, utilizing

SGD as the optimizer. A polynomial scheduler was employed to adjust the learning rate throughout the training process.

## Quantitative Results

Table 1 presents the Dice scores for our trained segmentation model on the baseline dataset. As shown, incorporating second-stage models enhances accuracy by focusing on the regions with inaccurate segmentation from the previous stage.

| **Stages** | **Models** | **Dice Score on Validation Set** |
|---|---|---|
| Stage 1 | DynUnet | 66.57 |
| Stage 2 | SegResNet | 67.10 |
| Stage 2 | SwinUNETR | 67.25 |
| Stage 2 | UNet | 66.88 |

Table 1: Quantitative results of trained models.

## Qualitative Results

Figure 1 illustrates representative examples of the segmentation results on both FDG and PSMA PET/CT images. These examples highlight the model's ability to accurately delineate tumor lesions while minimizing false positives in regions with physiologically high tracer uptake, such as the brain, heart, and kidneys.

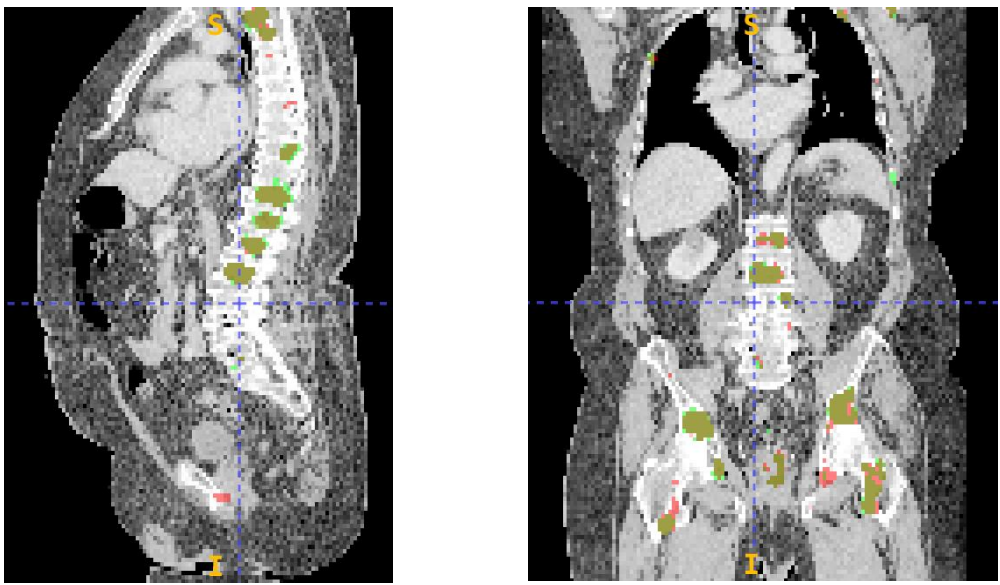

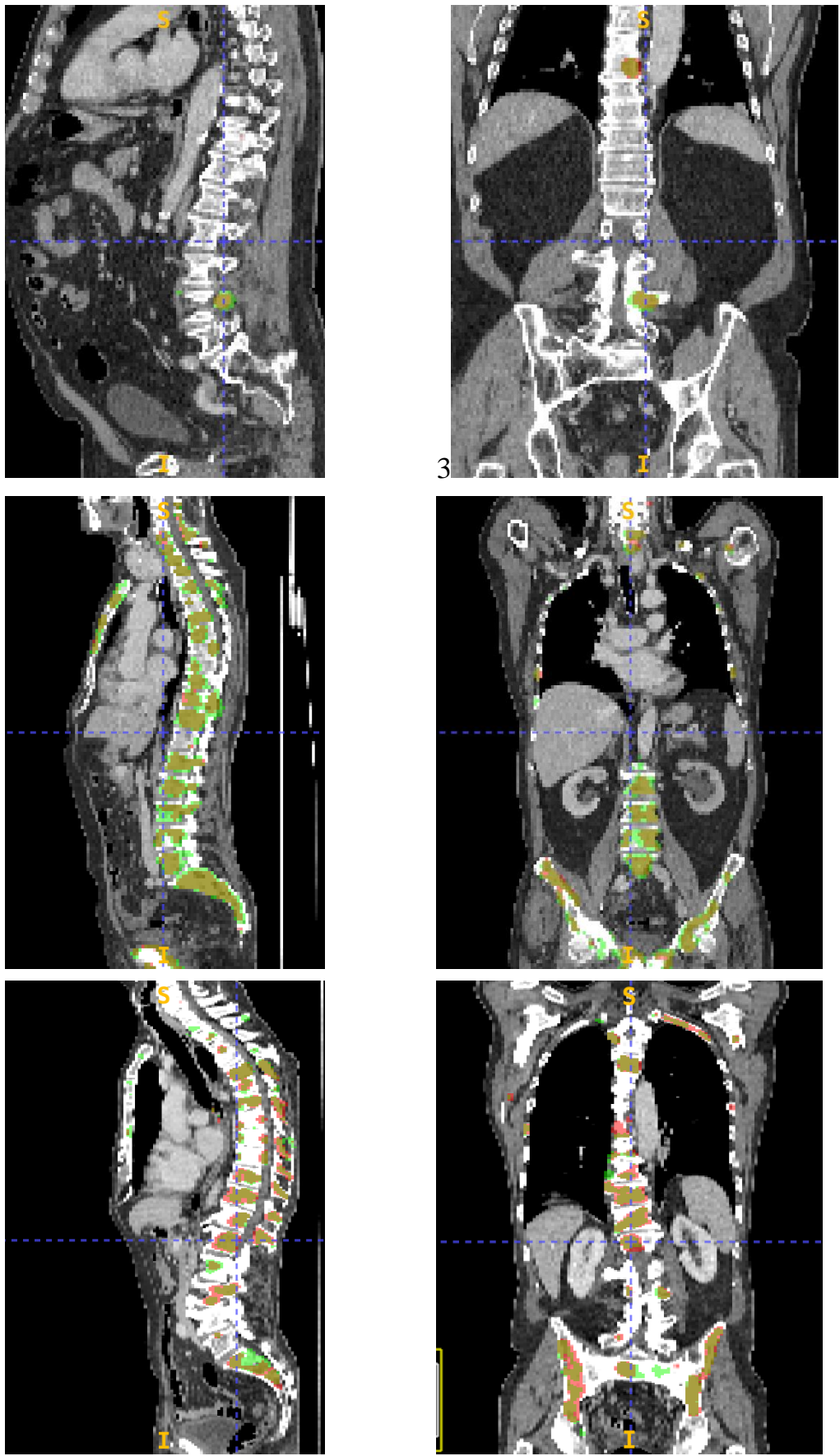

**Figure 1:** Visualization of some cases. The green regions are ground-truth. The red regions are the model predictions. Olive regions are the true positive area.

## Conclusion

In conclusion, this study successfully implemented a two-stage deep learning approach for automated lesion segmentation in PET/CT images as part of the AutoPET III challenge. By utilizing a DynUNet model for initial coarse segmentation and refining the results with an ensemble of SwinUNETR, SegResNet, and UNet models, the method demonstrated improved segmentation accuracy. The use of extensive data preprocessing, normalization, and data augmentation techniques further contributed to the robustness of the model. The results highlight the effectiveness of multi-stage segmentation and model ensembling in enhancing performance, particularly in addressing the complexities of multitracer multicenter datasets.

# References

[1] Gatidis S, Kuestner T. A whole-body FDG-PET/CT dataset with manually annotated tumor lesions (FDG-PET-CT-Lesions) [Dataset]. The Cancer Imaging Archive, 2022. DOI: 10.7937/gkr0-xv29

[2] Ingrisch, M., Dexl, J., Jeblick, K., Cyran, C., Gatidis, S., & Kuestner, T. (2024). Automated Lesion Segmentation in Whole-Body PET/CT - Multitracer Multicenter generalization. 27th International Conference on Medical Image Computing and Computer Assisted Intervention (MICCAI 2024). Zenodo. https://doi.org/10.5281/zenodo.10990932

[3] Jeblick, K., et al. A whole-body PSMA-PET/CT dataset with manually annotated tumor lesions (PSMA-PET-CT-Lesions) (Version 1) [Dataset]. The Cancer Imaging Archive, 2024. DOI: 10.7937/r7ep-3x37

[4] Isensee, F., Jäger, P.F., Kohl, S.A., Petersen, J., Maier-Hein, K.H.: Automated design of deep learning methods for biomedical image segmentation. arXiv preprint arXiv:1904.08128 (2019)

[5] Hatamizadeh, D. Swin UNETR: Swin Transformers for Semantic Segmentation of Brain Tumors in MRI Images. In Brainlesion: Glioma, Multiple Sclerosis, Stroke and Traumatic Brain Injuries 2022 (pp. 272–284). Springer International Publishing.

[6] Myronenko, A. 3D MRI Brain Tumor Segmentation Using Autoencoder Regularization. In Brainlesion: Glioma, Multiple Sclerosis, Stroke and Traumatic Brain Injuries 2019 (pp. 311–320). Springer International Publishing.

[7] Kerfoot, J. Left-Ventricle Quantification Using Residual U-Net. In Statistical Atlases and Computational Models of the Heart. Atrial Segmentation and LV Quantification Challenges 2019 (pp. 371–380). Springer International Publishing.